\documentclass{aa}
\usepackage{graphicx}
\usepackage{amsmath}
\usepackage{amssymb}
\usepackage{natbib}
\bibpunct{(}{)}{;}{a}{}{,} % to follow the A&A style
\begin{document}

\title{Is Arcturus a well-understood K giant?}
\subtitle{Test of model atmospheres and potential companion detection \\
by near-infrared interferometry}

\author{
T. Verhoelst \inst{1,3} \and P. J. Bord\'e \inst{2} \fnmsep
\thanks{Michelson Postdoctoral Fellow} \and G. Perrin \inst{3}
\and L. Decin \inst{1} \fnmsep \thanks{Postdoctoral Fellow of
the Fund for Scientific Research, Flanders} \and K. Eriksson
\inst{4} \and S. T. Ridgway \inst{3,5} \and P. A. Schuller
\inst{2} \and \\ W. A. Traub \inst{2} \and R. Millan-Gabet \inst{5}
\and M. G. Lacasse \inst{2} \and C. Waelkens \inst{1}
}

\institute{
Instituut voor Sterrenkunde, K.U. Leuven, Celestijnenlaan 200B, B-3001
Leuven, Belgium
\and Harvard-Smithsonian Center for Astrophysics, 60 Garden Street,
Cambridge, MA 02138, USA
\and Observatoire de Paris-Meudon, LESIA, 5 place Jules Janssen, 92195
Meudon, France
\and Institute for Astronomy and Space Physics, Box 515, 75120
Uppsala, Sweden 
\and National Optical Astronomy Observatories, PO Box 26732, Tucson,
AZ 85726, USA
\and Caltech/Michelson Science Center, Pasadena, CA 91125, USA
}

\date{Received 12 November 2004 / Accepted 23 January 2005}

\abstract{
  We present near-IR interferometric measurements of the K1.5 giant
  Arcturus ($\alpha$~Bootis), obtained at the IOTA interferometer with
  the FLUOR instrument, in four narrow filters with central
  wavelengths ranging from 2.03\,$\mu$m to 2.39\,$\mu$m. These
  observations were expected to allow us to quantify the wavelength
  dependence of the diameter of a typical K giant. They are compared
  to predictions from both plane-parallel and spherical model
  atmospheres. Unexpectedly, neither can explain the observed
  visibilities. We show how these data suggest the presence of a
  companion, in accordance with the Hipparcos data on this star, and
  discuss this solution with respect to Arcturus' single star status.

\keywords{Techniques: interferometric -- Stars: individual: \object{Arcturus} -- 
Stars: atmospheres -- Stars: binaries: general}
}

\titlerunning{Is Arcturus a well-understood K giant?}
\authorrunning{T. Verhoelst et al.}
\maketitle

%%%%%%%%%%%%%%%%%%%%%%%%%%%%%%%%%%%%%%%%%%%%%%%%%%%%%%%%%%%%%%%%%%%%%%

\section{Introduction}
\label{sec:intro}

Early K giants are often used as calibration sources for photometry,
spectroscopy and interferometry in the (near) IR because they offer a
good compromise between brightness at these wavelengths and
compactness of the atmosphere. The latter quality guarantees the
absence of exotic behaviour in both the line/band formation
(deviations from Local Thermodynamic Equilibrium (LTE), complicated
temperature distributions, ...) and in spatial structure.

One such very popular star is Arcturus (K1.5-2III, \object{$\alpha$\,Bootis}).
This star has been a stellar standard for many decades, not only in the
infrared for spectroscopic/photometric work, \citep[e.g. as an ISO-SWS
calibrator,][]{Decin2003d} but also as an IAU radial velocity standard
\citep[e.g.][]{Pearce1955}. Although its angular diameter of about
$20.20\pm0.08$\,mas \citep{Perrin98} implies that it is too
resolved by most interferometric configurations to be used as a
calibrator (Sect.~\ref{sec:calib}), it provides an excellent
opportunity to study the deviations from a non-wavelength-dependent
uniform-disk model (which is usually used to model the calibration
sources, a step necessary in the calibration process), and hence to
investigate the need for sophisticated models to calibrate
interferometric measurements if high-accuracy visibilities are
sought. Such a detailed study of Arcturus was already performed in the
visible wavelength regime by \cite{Quirrenbach96}, who found good
agreement with theoretical wavelength-dependent limb-darkening
predictions.  In this paper, we
investigate the near-IR part of the spectrum.
 
We start out by discussing the interferometric calibration process
(Sect.~\ref{sec:calib}). Then we present the data obtained on Arcturus
(Sect.~\ref{sec:obs}), after which we present different atmosphere models
with which one can interpret the data (Sect.~\ref{sec:interpret}).

Surprisingly, some data points are not at all consistent with the
proposed models, and seem to suggest a binary nature for Arcturus. We
discuss this solution and confront it with our current knowledge of
this star in Sect.~\ref{sec:binary}. Finally, in Sect.~\ref{sec:conclusions},
we summarize our results and the open questions.
 
%%%%%%%%%%%%%%%%%%%%%%%%%%%%%%%%%%%%%%%%%%%%%%%%%%%%%%%%%%%%%%%%%%%%%

\section{The interferometric calibration process: the need for well
  known calibrators}
\label{sec:calib}

Interferometric observations measure the spatial coherence of a given
source between two or more apertures. Any optical defect on the line
of sight such as atmospheric turbulence or in the instrument such as
polarization mismatches or dispersion will degrade spatial
coherence.

The classical solution to overcome this issue of decoherence is to
observe reference stars used as calibrator sources. The observed
degree of coherence measured on the calibrator (also called fringe
contrast) is compared to the value expected from prior knowledge of
the source characteristics. This defines the interferometric
efficiency (also called transfer function) of the instrument at the
time when the calibrator was observed. In single-mode interferometers
for which turbulent phase has been filtered out, the interferometric efficiency is
relatively stable. Yet, in order to monitor any instrumental change to
achieve the best accuracy, calibrator observations are interleaved
with science target observations.

In an ideal world, a calibrator should be point-like in order to yield
an expected visibility of 100\,\% to within an excellent
approximation.  However, if we receive non-zero flux from an
object then it must have a finite angular diameter. Furthermore, the
requirement of IR brightness and the more or less Planckian energy
distribution of stars, favours cool giants as IR interferometric
calibrators. Consequently in practice, calibrators are always
slightly resolved albeit with large visibilities. But an accurate
visibility estimate requires an accurate diameter estimate. Some
sources have been measured interferometrically or with the lunar
occultation technique. But for most calibrator sources an {\sl a
priori} estimate is required. It has to be based on spectroscopy,
photometry and modelling. Examples of such studies can be found in
\cite{Cohen96,Cohen99}, \cite{Borde02}, and \cite{Merand04} for the
near-infrared, or \cite{VanBoekel2004} for the mid-infrared.

These indirect techniques prove to provide excellent wide-band
diameter estimates with accuracies as good as a few percent, but higher
spectral resolution and/or an expected visibility of the calibrator
well below 100\,\% will require a better understanding of the
calibrator diameters. The observations presented here were made to
empirically quantify the wavelength-dependence of the diameter of an
early K giant and to test whether theoretical atmosphere models can be
used to compute the wavelength-dependent diameter of other
interferometric calibrators.

%%%%%%%%%%%%%%%%%%%%%%%%%%%%%%%%%%%%%%%%%%%%%%%%%%%%%%%%%%%%%%%%%%%%%%

\section{The observations}
\label{sec:obs}

Before presenting the new data, we briefly discuss the instrument, the
calibrator star and the data reduction strategy.

\subsection{The instrument}
The observations were performed in May 2002 at the IOTA
(Infrared-Optical Telescope Array) interferometer located at the
Smithsonian Institution's Whipple Observatory on Mount Hopkins,
Arizona \citep{traub98}. Several baselines of the IOTA have been used to
sample visibilities at different spatial frequencies. The data have
been acquired with FLUOR (Fiber Linked Unit for Optical Recombination)
in the version described by \cite{foresto98}. FLUOR is the precursor
of the now well-known VLTI/VINCI \citep{Kervella2000,Kervella2003}.
FLUOR has four outputs (two interferometric and two photometric) which
were focused on a Nicmos~3 array developed by \cite{millan-gabet96}
and operated with frame rates ranging from 500 up to 2000\,Hz.

Observations were carried out in narrow bands with filters specially
specified for molecular bands and the continuum region of cool stars
in K \citep[e.g.][]{Decin2000}. We will discuss in Sect.~\ref{sec:interpret}
which bands are present in the K band in the case of Arcturus and how
they are expected to influence the interferometric observations. The
narrow band filters transmissions are plotted in
Fig.~\ref{fig:spectrum}.  They are named K203, K215, K222 and K239
where the three digits characterize the central wavelength: 2.03,
2.15, 2.22 and $2.39\,\mu$m respectively.  The two
continuum filters K215 and K222 sample the maximum transmission region
of the K band.  The K203 (H$_{2}$O bands) and K239 (H$_{2}$O and CO
bands) sample the edges of the K band where stellar flux is attenuated
by the poorer transmission of the Earths atmosphere due to the
absorption by water vapor.  Any loss of coherence due to this water
vapour is taken into account by measuring the interferometric
efficiency in each band separatly.

\subsection{The calibrator star: HR\,5512}
Observations of Arcturus have been bracketed by observations of the
calibrator star \object{HR\,5512} (M5III) whose diameter is estimated to
$8.28\pm0.41$\,mas as explained in \cite{Perrin98}. Using an M5 giant
is of course rather risky in the sense that it probably also has a
slightly extended atmosphere with a wavelength dependent diameter. And
since it is partially resolved, the interferometric efficiency we used might be
biased. Theoretically, one can expect an atmospheric extension of
about 4--8\,\% for a late M giant \citep[e.g.][]{Wittkowski04}. At a
spatial frequency of 25\,arcsec$^{-1}$, this would induce a variation
with wavelength of the calibrator's visibility below
0.5\,\%. Nevertheless, this effect must be kept in mind when
interpreting the Arcturus data.

HR\,5512 does not appear in the
spectroscopic binary catalogues of \cite{Batten} and
\cite{Pourbaix}. {\sl Hipparcos} does not list it as a visual binary
either, though it does get the flag {\sl ``suspected binary''}. In
response, \cite{Mason1999} used speckle interferometry to possibly
resolve the system, but they found no companion. 

HR\,5512 is known to be a semi-regular (SR) variable star
\citep[e.g.][]{Percy1992} for which {\sl Hipparcos} found variations
with an amplitude of 0.1 mag and a period of only 6.3 days
\citep{Koen2002a}. However, more recently, samples of these {\sl
Hipparcos} short-period SR's have been the subject of other dedicated
photometric surveys, such as those performed by \cite{Koen2002b} and
\cite{Kerschbaum2001}, which could not confirm the presence of any
variability with a period below 35 days in any of their
targets. \cite{Kerschbaum2001} suggest some instrumental artefact on
the side of {\sl Hipparcos}. We are therefore confident that the
brightness of our calibrator HR\,5512 has not changed significantly
during the observing run. Furthermore, \cite{Koen2002b} find that in
most SR's, the brightness variations are primarily due to changes in
temperature, and not in diameter. This suggests that the diameter
variations of HR\,5512 during its pulsational cycle are well below the
2\,\% limit of the constant temperature scenario and hence do not affect
our science observations.

\subsection{Data reduction}
Fringe contrasts have been derived with the procedure explained in
\cite{FLUOR}. The bias in visibility estimates due to photon noise has
been removed following \cite{perrin2003a}. Turbulent corrugations of
the wavefront are cleaned by the single-mode fibers of FLUOR except
for the differential piston between two
apertures. \cite{perrinridgway2004b} have shown that the
piston-induced bias is smaller than the 0.1\,\% level for typical
fringe frequencies of a few hundred Hz as used in FLUOR. The expected
visibility of the calibrator is computed at the time it was observed.
The interferometric efficiency is then interpolated at the time when
Arcturus has been observed. Division of the fringe contrast of
Arcturus by the interpolated interferometric efficiency provides the
final visibility estimate. Correlations in fringe contrast and
transfer function estimates are taken into account in the computation
of error bars. The whole calibration procedure has been published in
\cite{perrin2003b}. The accuracy of visibility estimates measured with
the FLUOR setup is usually on the order of 1\,\% for most sources and
can be as good as 0.2--0.4\,\% \citep{perrin2003b,perrinridgway2004}.
In other words, all known biases are much smaller than the effects
reported in this paper and error bars are well estimated.

Table~\ref{table:obsdata} lists all measured visibilities in May 2002,
grouped per filter. 

\begin{table}[ht]
\begin{center}
\caption{All measured squared visibilities of the 2002 run, grouped per filter.
PA is the position angle (counted from North to East) of the interferometric
baseline projected onto the sky.}
\vspace{2ex}
\begin{tabular}{cccccc}
\hline
\hline
Date  & Filter & spat. freq.     & PA    & $V^2$ & $\sigma(V^2)$ \\
(MJD) &        & (arcsec$^{-1}$) & (deg) &       &               \\
\hline
52427 & K203  & 37.70 & 128.44 & 0.1656 & 0.0054 \\    
52429 & K203  & 37.27 & 122.93 & 0.1885 & 0.0050 \\    
52429 & K203  & 37.32 & 123.35 & 0.1901 & 0.0050 \\      
52430 & K203  & 26.68 & 122.37 & 0.4030 & 0.0117 \\      
52430 & K203  & 26.69 & 122.63 & 0.3961 & 0.0100 \\      
52431 & K203  & 37.62 & 131.03 & 0.1688 & 0.0048 \\   
\hline
52427 & K215  & 35.01 & 121.57 & 0.2347 & 0.0052 \\      
52429 & K215  & 35.25 & 123.75 & 0.2167 & 0.0055 \\      
52430 & K215  & 25.18 & 122.97 & 0.4533 & 0.0071 \\      
52431 & K215  & 35.46 & 131.41 & 0.2288 & 0.0054 \\  
\hline
52427 & K222  & 34.02 & 122.02 & 0.2581 & 0.0055 \\      
52429 & K222  & 34.25 & 124.28 & 0.2568 & 0.0057 \\      
52429 & K222  & 34.30 & 124.74 & 0.2535 & 0.0056 \\      
52430 & K222  & 24.44 & 123.38 & 0.5336 & 0.0108 \\      
52431 & K222  & 34.36 & 131.70 & 0.2522 & 0.0058 \\  
\hline   
52425 & K239  & 42.09 &  93.42 & 0.0867 & 0.0054 \\      
52427 & K239  & 31.08 & 116.61 & 0.2937 & 0.0083 \\      
52429 & K239  & 32.00 & 125.74 & 0.2861 & 0.0111 \\      
52430 & K239  & 22.74 & 124.06 & 0.5719 & 0.0099 \\      
52431 & K239  & 31.94 & 132.04 & 0.2886 & 0.0082 \\       
\hline
\end{tabular}
\label{table:obsdata}
\end{center}
\end{table}

%%%%%%%%%%%%%%%%%%%%%%%%%%%%%%%%%%%%%%%%%%%%%%%%%%%%%%%%%%%%%%%%%%

\section{Interpretation of the data: comparison with theoretical
  atmosphere models}
\label{sec:interpret}

Typical of present-day optical interferometric data is that they do not allow
an inverse Fourier Transform (to obtain an image) but only offer
information when they are compared to a model of the source intensity
distribution on the sky. This implies that one needs to know in
advance, through previous (non-interferometric) studies, which type of
model to choose. For Arcturus, a first step could be the fitting
of a uniform disk model for each individual filter which yields the
apparent diameters at the different wavelengths. These values can then
be compared to theoretical predictions. While this strategy is
satisfactory for the study of stars with very extended atmospheres for
which accurate model atmospheres are not yet available and for which
the observed change of size with wavelength is large, it is not
sophisticated enough for Arcturus, for which we expect diameter
variations with wavelength of only a few percent \citep{Decin2003d}. Hence, we would
like to use a fully self-consistent theoretical model atmosphere and
compute from this a wavelength-dependent, limb-darkened synthetic
source intensity distribution on the sky.

\subsection{Arcturus atmosphere models and diameter determination}

In this section, we present our dedicated {\sc marcs} model
atmospheres and the resulting spatial intensity profiles. These are
then converted into synthetic visibilities and compared with the
observations. 

\subsubsection{The atmosphere models}

Arcturus was used as a primary calibrator for the calibration of
the ISO-SWS (Infrared Space Observatory Short Wavelength Spectrometer)
\citep{Decin2000} and was for these purposes modelled using the {\sc
marcs}-code \citep[][and references therein]{Plez1992}. This code
is aimed at the modelling of atmospheres of cool (giant) stars,
allowing both plane-parallel (PP) and spherical (SPH)
geometries. Basic assumptions underlying the models are the following:
LTE, hydrostatic equilibrium and conservation of energy for radiative
and convective flux. The radiative transfer equation is solved using
an ALI (Approximate Lambda Iteration) method \citep{Nordlund} with the
OS (Opacity Sampling) technique as the statistical way of treating
spectral lines.

\cite{Schmid-Burgk} and \cite{Scholz85} found that the PP
approximation was no longer valid in cool giants when a measure of the
extension of the atmosphere, $d$, defined to be 
\begin{equation} \label{eq:def_ext}
d \equiv \frac{R_{\tau_\mathrm{Ross} = 10^{-5}} -
R_{\tau_\mathrm{Ross} = 1}}{R_{\tau_\mathrm{Ross} = 1}}
\end{equation}
with R the linear radius and $\tau_\mathrm{Ross}$ the Rosseland
optical depth, was larger than 0.05.

Recent studies of the role of sphericity in oxygen-rich cool stars
include those of \cite{Scholz85}, \cite{Bessell89} and
\cite{Plez1992}. Compared to PP models, the radiation
field in a SPH model becomes diluted in the upper photospheric
layers. This usually leads to a decrease in temperature of the surface
layers which can influence significantly the molecular opacity
\citep[][p.~650]{Mihalas78}.

SPH models have to deal with the added complexity of an explicit
radial dependence of all variables, plus a more complex equation of
radiative transport,  solved for different rays toward the observer:
\begin{equation}
\frac{\mu}{\kappa_{\nu} \rho} \frac{\partial I_{\nu}}{\partial r} +
\frac{1-\mu^2}{\kappa_{\nu} \rho r} \frac{\partial I_{\nu}}{\partial
\mu}  =  S_\nu - I_\nu
\end{equation}
with $\mu = \cos{\theta}$ the cosine of the viewing angle,
$\kappa_{\nu}$ the monochromatic extinction coefficient per unit
mass, $\rho$ the density, $I_\nu$ the specific intensity and $S_\nu$
the source function. Computationally, this equation is solved
by using the single-ray approximation \citep{Nordlund} for a set of
parallel rays, intersecting the atmosphere at different values for the
impact parameter. In computer time, the difference between
PP and SPH models is usually less than a factor of three although
convergence is not guaranteed in the latter case.

\begin{figure}
  \resizebox{\hsize}{!}{\includegraphics{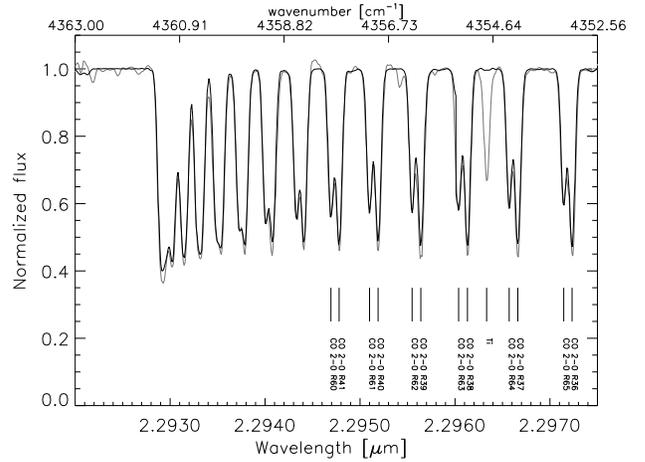}}
  \caption{FTS spectra (summer and winter, in grey) compared to our
  synthesized spectrum (black) at the  $^{12}$CO
  $2-0$ bandhead near 2.29~$\mu$m. The good match confirmes the
  quality of our model in the FLUOR wavelength region. The discrepancy
  in the Titanium line is due to a poorly known $\log{gf}$ value. 
  }
  \label{fig:FTS}
\end{figure}

 The high quality of the fit between the resulting synthetic spectrum and
both the observed ISO-SWS spectrum and high-resolution FTS spectra presented
in \cite{Decin2003b}, suggests that these hydrostatic models offer a good
representation of the true atmosphere of Arcturus. Since this
comparison was limited to the ISO-SWS wavelength region ($2.38 - 45 \mu$m), we
searched the full Arcturus FTS spectral atlas \citep{Hinkle1995} in
the FLUOR bandpasses for peculiar spectral features. Lines are sparse
(mainly CN and some atomic lines) and well spaced up to the $^{12}$CO
$2-0$ bandhead at 2.29~$\mu$m. A comparison between the FTS spectra
(summer and winter) and a spectrum synthesized from our atmosphere
model at this bandhead is shown in Fig.~\ref{fig:FTS}. Clearly the
match is excellent, confirming the quality of our model. 

%
%_______________________________________________Fig. FTS
%

\begin{figure}
  \resizebox{\hsize}{!}{\includegraphics{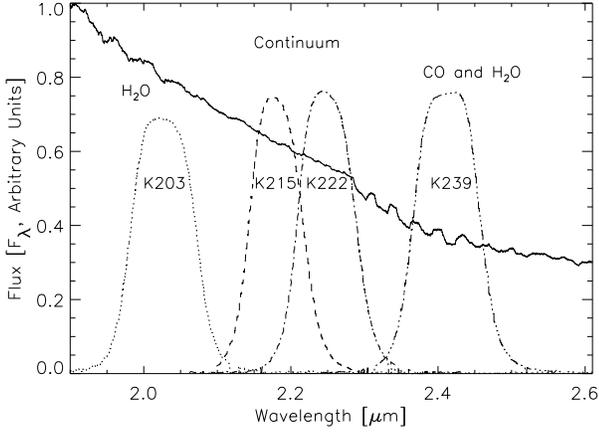}}
  \caption{
   The transmission curves of the 4 narrow band filters used in the
   FLUOR instrument on IOTA together with a synthetic K-band spectrum
   of Arcturus. For cool late-type stars, the filter centered at
   2.03\,$\mu$m probes a H$_2$O band in the spectrum (not present in
   Arcturus). For a K giant, the filters at 2.15 and 2.22\,$\mu$m both
   probe the continuum, and the 4th filter at 2.39\,$\mu$m covers the
   CO first overtone band (and possibly also H$_2$O).
  }
  \label{fig:spectrum}
\end{figure}

From the quality of the fit between synthetic and observed spectra, we
find little reason to prefer a SPH model over a PP one, suggesting
that the atmosphere is quite compact (indeed, $d = 0.02$ for the
model). Nevertheless, we computed both PP and SPH atmospheres for a
grid around the stellar parameters determined by \cite{Decin2003d} for
Arcturus (listed in Table~\ref{table:params}), since the effects of
sphericity may still be detectable in the intensity profile on the sky
and hence also in the visibilities.  From these models, we derived
a spatial intensity profile for each OS wavelength point in the K
band. The full OS wavelength grid of our model contains about 150,000
wavelength points, with a resolution of R$\sim$20,000 at
2~$\mu$m. These points are chosen in such a way as to accurately
sample the total opacity as a function of wavelength and is based on
extensive atomic/molecular linelists and continuum opacity
sources. The full K band spectrum of this model is shown in
Fig.~\ref{fig:spectrum}, together with the FLUOR filter profiles.

\begin{table}[ht]
 \begin{center}
 \caption{
 The stellar parameters from \cite{Decin2003b} and references therein, and the limits
 of the grid in which we searched for an optimal fit to the
 interferometric data ($\xi_{\rm{t}}$ is the microturbulent velocity).
 }
 \vspace{2ex}
 \begin{tabular}{lrrr}
 \hline
 \hline
 Parameter & Value & Lower limit & Upper limit \\
 \hline
 $T_\mathrm{eff}$ (K)            &  $4320\pm140$    & 4250    & 4500    \\
 $\log g \: (\mathrm{cm/s}^2)$   &  $1.50\pm0.15$  &  1.00   & 2.00    \\
 $[\rm{Fe/H}]$                   &  $-0.5\pm0.20$  &  $-1.0$ & $0.0$   \\
 $\xi_{\rm{t}}$ (km/s)           &  $1.7\pm0.3$    &     &         \\
 \hline
 \end{tabular}
 \label{table:params}
 \end{center}
\end{table}

\begin{figure}
  \resizebox{\hsize}{!}{\includegraphics{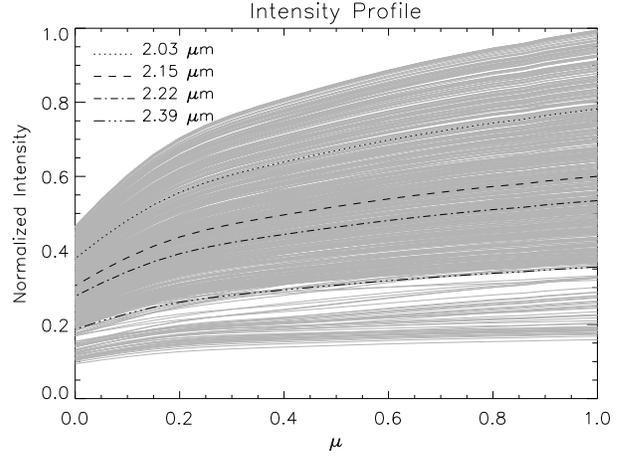}}
  \caption{
  Intensity profiles for a plane parallel model at T$_{\rm{eff}} =
  4250\,K$ and $\log{g} = 1.0$, normalized to the maximum intensity
  in the K-band at the center of the stellar disk. Every gray curve
  was computed for an OS wavelength within the K band. Clearly there
  is a wide range of shapes over the stellar disk, but integration
  over the bandpasses reduces the differences. Because the model
  is semi-infinite for all viewing angles, $\mu=0$ corresponds to a
  singularity.
  }
  \label{fig:PPprofiles}
\end{figure}

\begin{figure}
  \resizebox{\hsize}{!}{\includegraphics{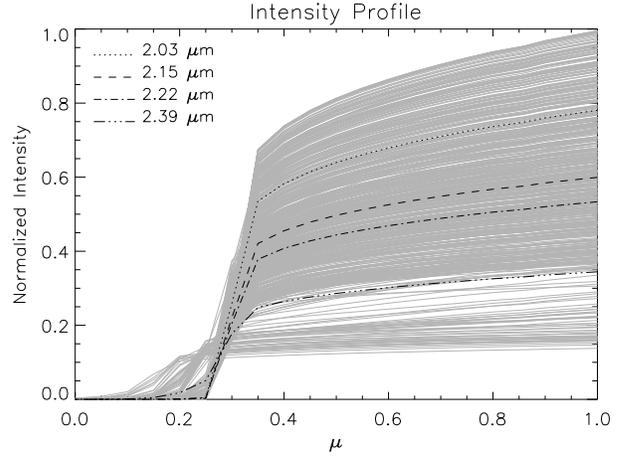}}
  \caption{
  Intensity profiles for a spherical model at T$_{\rm{eff}} = 4250\,K$
  and $\log{g} = 1.0$, normalized to the maximum intensity in the K-band
  at the center in the stellar disk.  Every gray curve was computed for an OS
  wavelength within the K band. We now see even more variety than in
  Fig.~\ref{fig:PPprofiles}, mainly toward the stellar limb. The 2.39\,$\mu$m
  filter differs significantly from the others in this outer
  region$^1$.
  }
  \label{fig:SPHprofiles}
\end{figure}

The resulting variety in intensity profiles (with wavelength) is
largest for the model with stellar parameters $T_\mathrm{eff} =
4250$\,K and $\log{g} = 1.00$ and is shown in
Fig.~\ref{fig:PPprofiles} and Fig.~\ref{fig:SPHprofiles}. To improve
display of the profile at the limb, we plot intensity as a function of
$\mu = (1-r^2/R_{\tau_{\mathrm{Ross}} = 10^{-7}}^2)^{1/2}$, where $r$ is the
projected linear distance from the center of the stellar disk.  The
FLUOR-bandpass-integrated profiles are overplotted in black. Even in
the PP geometry there is a noticeable variety in profiles, but through
the integration over the bandpasses we lose most of that
information. In the SPH geometry, the variety in profiles is much
stronger and even after integration differences do remain, mainly in
the outer regions of the disk in the CO and H$_2$O probing
filter\footnote{
   We remark that high spectral resolution interferometric
   observations, e.g. with VLTI/AMBER \citep{Petrov2000} will resolve
   the individual spectral lines visible between
   $\mu=0.2~\mathrm{and}~0.0$. The difference in visibility curve
   between such a spectral line and the continuum will be much
   stronger than that between the different FLUOR filters discussed here.}
In this case, this is due only to CO, since the model predicts no lines
of water vapor (the photospheric temperature is too high for this molecule
to survive).

Because of circular symmetry, these intensity profiles can now be
Hankel-transformed into visibility curves \citep{Hanbury-Brown74}
\begin{equation} \label{eq:LD}
V(x) = \frac{\int_0^1 I(\mu) \, J_0(\pi x \, \phi \sqrt{1-\mu^2}) \,
\mu \, \mathrm{d} \mu}{\int_0^1 I(\mu) \, \mu \, \mathrm{d} \mu},
\end{equation}
where $x = B/\lambda$ is the spatial frequency (arcsec$^{-1}$)
 with $B$ the projected baseline and $\lambda$ the wavelength, $I$
the intensity profile, $J_0$ the zeroth-order Bessel function of
the first kind, and $\phi$ the angular diameter. Here, the angular
diameter corresponds to the image size, i.e. the outermost point of
the model ($\tau_{\mathrm{Ross}} = 10^{-7}$). It is the
only free parameter of the model and is determined by fitting the
theoretical curves to the observed data. The results of this fitting
are presented in the next section.

\subsubsection{Diameter definitions}
\label{sec:diameters}

Comparison between this image size and the different types of diameters
generally used in the literature is not trivial. In the case of a
uniform disk (UD), the visibility is given by
\begin{equation} \label{eq:UD}
V(x) = \frac{2J_1(\pi x \phi)}{\pi x \phi}
\end{equation}
with $x$ and $\phi$ as in Eq.\,\ref{eq:LD}, and $J_1$ the first-order
Bessel function of the first kind. Fitting a UD model to the data,
we find the set of angular diameters listed in Table~\ref{table:UD_diam}.
These values are not compatible with a single
  wavelength-independent UD, and in addition the fit
quality appears very poor in most cases (column 3 of
Table~\ref{table:UD_diam}), thus demonstrating the
need for atmosphere models when precisions of $\approx 1\,\%$ are reached on
angular diameters.

\begin{table}[ht]
\caption{Uniform disk (UD) angular diameters derived by fitting Eq.~\ref{eq:UD}
to the narrow-band data taken in respective filters or altogether. The third
column gives the reduced chi-square of the fit (chi-square per degree of freedom),
denoted $\chi^{2}_{\mathrm{r}}$.
}
\label{table:UD_diam}
\vspace{1ex}
\begin{center}
\begin{tabular}{lccc}
\hline
\hline
filter & $\phi$ (UD, mas)   & $\chi^{2}_{\mathrm{r}}$ \\
\hline
K203   & $20.77 \pm 0.18$ & 9.20 \\
K215   & $20.91 \pm 0.15$ & 8.26 \\
K222   & $20.44 \pm 0.16$ & 0.40 \\
K239   & $21.12 \pm 0.25$ & 2.01 \\
all    & $20.72 \pm 0.16$ & 6.14 \\
\hline
\end{tabular}
\end{center}
\end{table} 

Alternatively, one may be interested in the limb-darkened (LD)
diameter. This diameter is the actual physical diameter of the star,
but is only well defined for stars with a compact atmosphere (i.e. in
case of a high surface gravity). Since the outermost radial point of
our atmosphere model ($R_{\tau_{\mathrm{Ross}} = 10^{-7}}$) is
supposed to correspond to the physical boundary of the star, the LD
diameter is actually the image size ($\phi$) in the case of a SPH
model (cf. Eq.~3 and the definition of $\mu$).  This is not correct for
the PP models because these are semi-infinite for all viewing angles
(with a singularity at $\mu = 0$), therefore the value of the
intensity profile at $\mu =0$ is an extrapolation from the last
calculated $\mu$ point. This last calculated $\mu$ point depends on
wavelength, because different wavelengths come with different numbers
of rays in the model. Figures~\ref{fig:PPprofiles} and
\ref{fig:SPHprofiles} demonstrate this difference between PP and SPH
models. Consequently, the image size listed in Column~2 of
Table~\ref{table:diameters} does not correspond to
$R_{\tau_{\mathrm{Ross}} = 10^{-7}}$ and can not be converted into a
$\tau_{\mathrm{Ross}}=1$ diameter as is done for the SPH models below. 

A physically relevant and well defined diameter is the one
corresponding to the $\tau_\mathrm{Ross} = 1$ layer. It can be derived
from the image size (of a SPH model) if one knows the ratio of the
outermost radial point in the model to the $\tau_\mathrm{Ross} = 1$
diameter. This ratio is similar to what is called the ``extension of
the atmosphere'', $d$, as defined in Eq.\,\ref{eq:def_ext}, but for
the models presented here, the outermost point corresponds to
$\tau_\mathrm{Ross} = 10^{-7}$. This $\phi_{\tau_\mathrm{Ross}=1}$ is also presented in Table~\ref{table:diameters}.

\begin{table}[ht]
\caption{Angular diameter determinations with plane-parallel and spherical
  {\sc marcs} models. Only a very small improvement on the chi-square can
  be achieved by increasing the extension of the atmosphere of the model (SPH model
  with $\log{g}=1.0$), and such a low surface gravity cannot be reconciled
  with the spectral features. 
}
\label{table:diameters}
\vspace{1ex}
\begin{center}
\begin{tabular}{lccc}
\hline
\hline
model                           & PP, 4250K      & SPH, 4250K       & SPH, 4250K       \\
                                & $\log g=1.5$   & $\log g=1.0$     & $\log g=1.5$     \\
\hline
$\phi$ (mas)                    & $21.32\pm0.19$ & $22.37 \pm 0.19$ & $21.84 \pm 0.21$ \\
$\phi_{\tau_\mathrm{Ross}=1}$   & NA             & $21.16 \pm 0.18$ & $21.19 \pm 0.20$ \\
$d$ (Eq.\,\ref{eq:def_ext}$^*$) & 0.030          & 0.057            & 0.030            \\
$\chi^{2}_{\mathrm{r}}$         & 5.63           & 5.38             & 5.51             \\

\hline
\multicolumn{4}{l}{$^*$ out to $\tau_\mathrm{Ross} = 10^{-7}$}\\
\end{tabular}
\end{center}
\end{table}

\begin{figure}
  \resizebox{\hsize}{!}{\includegraphics{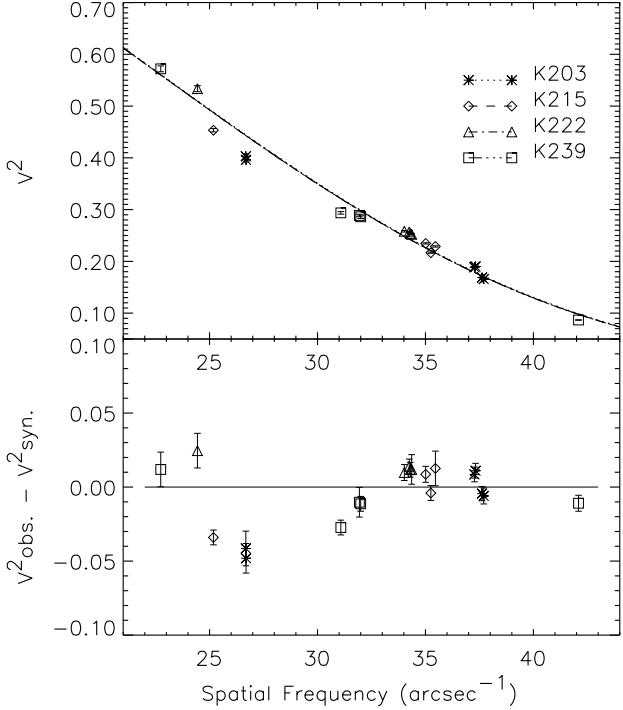}}
  \caption{
  A comparison between the FLUOR visibility measurements of Arcturus and a
  plane-parallel {\sc MARCS} visibility model ($T_\mathrm{eff} = 4250$\,K,
  $\log{g} = 1.50$; $\chi^2_{\mathrm{r}} = 5.63$ for $ \phi = 21.32
  \pm 0.19$\,mas). The four theoretical curves coincide, while there
  is a significant dispersion of the data points.
}
  \label{fig:pp}
\end{figure}

\begin{figure}
  \resizebox{\hsize}{!}{\includegraphics{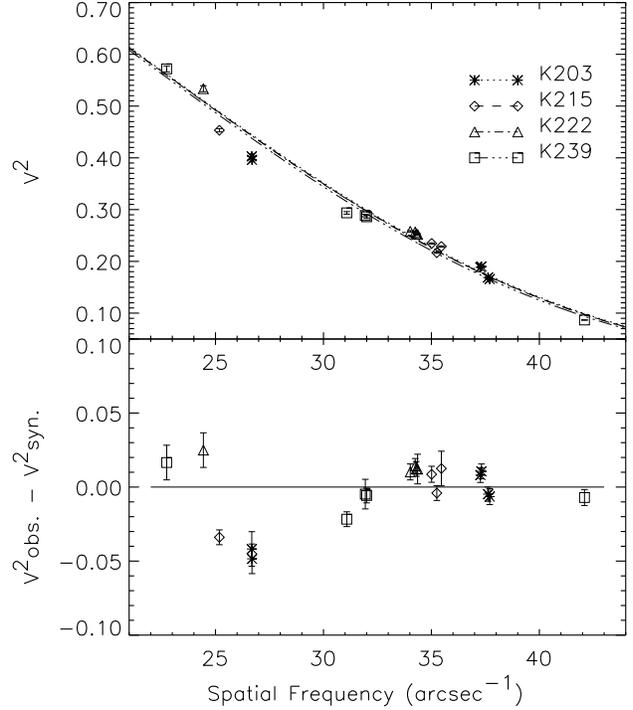}}
  \caption{
  For the spherical model with slightly lower surface gravity ($T_\mathrm{eff} = 4250$\,K,
  $\log{g} = 1.00$; $\chi^2_{\mathrm{r}} = 5.38$ for $ \phi = 22.37 \pm 0.19$\,mas),
  the visibility curve in the CO-probing filter no longer coincides with
  the others, corresponding to a slightly larger diameter, but it
  still cannot match the measured visibilities.
  }
  \label{fig:sph}
\end{figure}

Figures~\ref{fig:pp} and \ref{fig:sph} show a comparison of our
theoretical visibility curves with the observed narrow-band visibilities.
For the PP model (Fig.~\ref{fig:pp}), all four visibility curves coincide.
The best-fit angular diameter for $T_\mathrm{eff} = 4250$\,K, $\log{g} = 1.50$
and $[\rm{Fe/H}] = -0.50$, is $\phi = 21.32 \pm 0.19$\,mas.  The chi-square per
degree of freedom or reduced chi-square is then $\chi^2_\mathrm{r} = 5.63$. 

For the SPH model with the same stellar parameters (Fig.~\ref{fig:sph}), the
filter centered within the CO band around 2.39\,$\mu$m shows a slightly larger
star, though that is barely visible in the plot. We obtain $\phi = 21.84
\pm 0.21$\,mas for $T_\mathrm{eff} = 4250$\,K, $\log{g} = 1.50$,
$[\rm{Fe/H}] = -0.50$. However, the improvement is small since
$\chi^2_\mathrm{r} = 5.51$ for this model. Some improvement can be achieved by
reducing the surface gravity of the model (increasing the extension of the
atmosphere): this leads to a significantly different visibility curve for
the CO filter, but the improvement in terms of $\chi^2_\mathrm{r}$ is very
small (cf. Table\,\ref{table:diameters}), and such a low surface gravity cannot
be reconciled with the observed spectral features \citep{Decin2003b}. 

We remark that our newly derived diameters are slightly (but
significantly) larger than the LD diameters determined in K broadband
\citep[$20.91 \pm 0.08$,][]{Perrin98} and in the optical \citep[$21.0
\pm 0.2$ mas, ][]{Quirrenbach96}. The bias introduced by the use of a
single calibrator is taken into account in the error determination and
should thus not be responsible for this discrepancy. The diameter of
Arcturus was also estimated with spectro-photometric techniques
\citep[e.g.][]{Cohen99, Decin2003b, VanBoekel2004}, but uncertainties
on those diameters are generally much larger and therefore compatible
with both our results and those discussed above. The uncertainties on
the distance ($11.26 \pm 0.09$\,pc) derived from the parallax, the
surface gravity and the mass, make it impossible to obtain an accurate
estimate of the angular diameter which does not depend on observed
flux levels and direct (interferometric) measurements. The recently
derived empirical surface brightness relation by \cite{Kervella2004}
yields a diameter of $21.2\pm0.2$ (assuming a 0.02 mag error on the K
magnitude) which is well in agreement with our result.

\subsection{Discussion}
\label{sec:atmos_discussion}

To our surprise, neither PP nor SPH models can explain the observed
visibilities. Indeed, the probability to obtain $\chi^2_{\mathrm{r}}
\approx 5.6$ with 19 degrees of freedom is as low as $10^{-14}$,
and there appear to be systematics in the residuals. In the following,
we investigate possible sources of these discrepancies. 

 An important characteristic of the residuals is that they not
  only point at a problem with the wavelength dependence of the
  visibility, but also at a problem with the shape of the visibility
  curve which is not consistent with that of a limb-darkened disk: the
  large discrepancy at a low spatial frequency (25~arcsec$^{-1}$)
  suggests a structure at least a few times larger than the stellar disk.

\subsubsection{Calibration problems}

A first check that needs to be done concerns the calibrator source: as
discussed in Sect.~\ref{sec:obs}, the extension of the calibrator's
diameter should contribute at most only 0.5\,\% to the residuals at
25~arcsec$^{-1}$. It is however possible that HR\,5512 shows a more
complicated atmosphere/circumstellar environment, but some simple
calculus shows that ignoring this effect in the calibrator actually
leads to an underestimation of the variation in visibility with
wavelength in the science target: if the extension of the calibrator
is affecting the calibration, then the problem for Arcturus is
actually even larger than the few percent reported here. Only if we
assume the data points that fall below the fit (e.g. those at
26\,arcsec$^{-1}$) to be correct, can the extension of the calibrator
cause the other points to be overestimated. However, these data points
are not at all compatible with photometric diameter estimates of
Arcturus \citep[e.g. 20.8\,mas,][]{Decin2000}.  Moreover, in such a
scenario, we expect problems mostly in the wavelength dependence and
not in the shape of the visibility curve.

\begin{figure}
  \resizebox{\hsize}{!}{\includegraphics{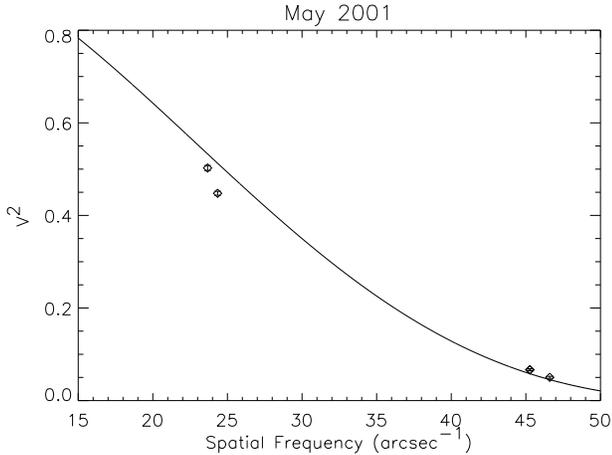}}
  \caption{
  Narrow-band data on Arcturus obtained in 2001 with different
  calibrators. The best fitting LD disk
  model (full line) shown here ($\chi^2_{\mathrm{r}} = 64.5$) leaves
  the same kind of residuals as seen in the 2002 data. 
  }
  \label{fig:2001data}
\end{figure}

Conclusive evidence against problems with the calibrator is provided
by a limited set of similar narrow-band FLUOR data on Arcturus from a
run in May 2001 (Table~\ref{table:2001}). Unlike the 2002 data
presented in this paper, these 2001 data were calibrated with
different calibrators, not including HR\,5512. They are presented in
Fig.\,\ref{fig:2001data} together with the best fit LD disk model. It
is clear that these data show very similar residuals.  We thus
conclude that the effects reported in this paper cannot be due to the
calibrator.

\begin{table}[ht]
\begin{center}
\caption{All measured squared visibilities of the 2001 run, grouped per filter.\label{table:2001}}
\vspace{2ex}
\begin{tabular}{cccccc}
\hline
\hline
Date  & Filter & spat. freq.     & PA    & $V^2$ & $\sigma(V^2)$ \\
(MJD) &        & (arcsec$^{-1}$) & (deg) &       &               \\
\hline
52043 & K222 &  23.67 &  98.30 &  0.5079 & 0.0054 \\ 
52047 & K222 &  45.27 &  95.51 &  0.0684 & 0.0021 \\
52043 & K215 &  24.34 &  95.47 &  0.4522 & 0.0044 \\
52045 & K215 &  45.27 &  89.02 &  0.0689 & 0.0020 \\
52047 & K215 &  46.60 &  93.07 &  0.0525 & 0.0022 \\
\hline
\end{tabular}
\end{center}
\end{table}

%
%_________________________________________ Fig. 2001 data
%

The largest deviations from the expected visibility where observed on
only one night (MJD=52430), and one might wonder whether observing
conditions showed any peculiarities that night. We found no sign of
this night being any different from the others of that run. Strong
piston often induces a broadening of the power spectrum of the affected
fringe scans, but this is not seen in the data presented here. 

\subsubsection{The photosphere of Arcturus}

Recently, \cite{Ryde} reported the presence of H$_2$O lines in a
high-resolution Texes spectrum of Arcturus, which were not present in
the {\sc marcs} model used for comparison (also used
here). \cite{Ryde} show how the presence of these lines can be
produced in the synthetic spectrum by imposing a slightly lower
temperature in the outer layers (log\,$\tau_{500} \leq -4 $) of the
model-atmosphere.  Although this new temperature distribution will
influence somewhat the location of the line-forming region, the fact
that the differences with the original {\sc marcs} model are only
visible in a high-resolution spectrum suggest that it should not
influence our low-resolution interferometric measurements.

 Still,
recent results on supergiant stars demonstrate that additional,
non-photospheric layers of water can have such a combination of
geometry and temperature that the extra absorption of stellar flux is
filled in by the emission of the outer regions of the layer, leaving
little trace of the significant additional water column density at
medium spectroscopic resolution \citep[e.g. ][]{Ohnaka2004}. This
additional water opacity might still be detectable through individual
lines in high-resolution spectra, such as the Texes spectrum and will
significantly influence the interferometric observations. However,
these additional water layers do create a {\sl pseudo} continuum,
affecting the spectro-photometric flux levels within the molecular
bands. While this effect might still be minimal for one given band,
the same temperature-geometry combination of the layer will result in
a strong signature in another band. For Arcturus, \cite{Decin2003b}
show that discrepancies between our {\sc marcs} model and the ISO
spectrum are not present in any water band between 2.38 and
12~$\mu$m. Moreover, the existence of non-photospheric molecular
layers is generally linked to the pulsation of the central star, and
is therefore unlikely in the case of Arcturus. 

We might also wonder whether the pressure scale height coming out of
the model is correct. It is known that this is not the case for cooler
and more luminous stars \citep{Perrin2004b}. However, fitting a model
with a larger atmospheric extension did not improve the $\chi^2$
significantly (see Table\,\ref{table:diameters}), and the decrease in
surface gravity required to obtain such an atmosphere is not
compatible with spectroscopic data. 

Furthermore, we must note that the
dispersion seen in the FLUOR data is not only a molecular-band
vs. continuum effect: the data obtained on the shortest baseline
(around 25\,arcsec$^{-1}$) show that even for the continuum filters, not all
points are consistent with a limb-darkened disk geometry. This feature
actually rules out the hypothesis of an under-estimation of the
extension of Arcturus' atmosphere: even if more pronounced than in the
SPH models, it could never influence the intensity profile in the
continuum to this extent.

Another hypothesis could be the presence of spots on the stellar
surface. However, this extra small scale structure would produce
oscillations in the visibility curve whose period must be larger than
the first null spatial frequency and hence it cannot explain our
data. If we believe these data points are correct (and we see no
reason not to), more exotic solutions should be explored.

We are of course aware of Arcturus' status as calibrator source for
many different instruments/techniques and of the research done on this
object in that framework but the following subsection will show that
some further investigation is nevertheless necessary to explain our
interferometric data and, for example, also the {\sl Hipparcos} data on
Arcturus.

%%%%%%%%%%%%%%%%%%%%%%%%%%%%%%%%%%%%%%%%%%%%%%%%%%%%%%%%%%%%%%%%%%%%%

\section{The binary hypothesis}
\label{sec:binary}

In this section, we investigate the possibility that the discrepancy
between the {\sc marcs} model and our interferometric data might be due
to the presence of a faint companion. Moreover, we discuss the
consistency of this hypothesis in the light of previously published works.

\subsection{Binary model}
\label{solution}

Let us denote by $V_1$ and $V_2$ the visibilities of Arcturus and its
companion considered as a single stars. The companion is assumed to be
unresolved ($\phi_2 = 0$), therefore $V_2 = 1$, and for $V_1$ we use
the limb-darkened model of Eq.~\ref{eq:LD}.   The contrast ratio
between the two stars, $r \equiv F_1/F_2$ with
$F$ the received flux, is assumed to be identical for all four
filters. This might be too crude an approximation if the spectral
types of the two stars differ significantly, but we show in
Sect.~\ref{sec:nature_companion} that this cannot be the case. We
neglect the motion of the companion during the 7 days of our
observation run, so that its separation $\rho$ and position angle
$\theta$ would remain constant. The squared modulus of the visibility
for the binary then reads
\begin{equation} \label{eq:v2_bin}
V^2 = \frac{r^2V_1^2+V_2^2+2rV_1V_2\cos[2\pi (B/\lambda) \rho
\cos(\theta-\theta_B)]}{(r+1)^2},
\end{equation}
where $(B,\theta_B)$ are the polar coordinates of the interferometric
baseline vector projected onto the sky. Note that for this equation to
be valid, the companion's fringe packet should be well overlapped with
the primary star's fringe packet. Since the narrow filters used here
guarantee about 55 fringes in the fringe packet, while the derived
separation between the 2 components (Table~\ref{tab:binary})
corresponds to only 7 fringes, Eq.~\ref{eq:v2_bin} can be used.
The parameters $\rho$, $\theta$, $r$ and $\phi_1$ are estimated by
minimizing a standard $\chi^2$ with the Levenberg-Marquardt procedure.
One should be particularly careful in the search for the global minimum
as trigonometric functions in Eq.~\ref{eq:v2_bin} cause the $\chi^2$
hypersurface to have many local minima. To overcome this problem,
we use a four-dimensional grid of initial guesses to run the minimization,
and keep the best $\chi^2$ of all. With this set of data, the estimation
is made more difficult by the fact that the visibility curves were not sampled
with the idea of looking for a binary (a continuous sampling at one
wavelength would have been much more efficient).

\begin{table}[ht]
\centering
\caption{Best-fit parameters with formal errors for the binary model.}
\vspace{2ex}
\begin{tabular}{lll}
\hline
\hline
                    & This work (K band)   & {\sl Hipparcos} (V band) \\
\hline
$\rho$              & $212.7 \pm 1.5$ mas  & $255 \pm 39$ mas   \\
$\theta$            & $157.6 \pm 1.7$ deg  & 198 deg            \\
$r$                 & $50.1 \pm 5.6$       & $21.5 \pm 5.9$     \\
$\Delta m$          & $4.25 \pm 0.12$      & $3.33 \pm 0.31$    \\
$\phi_1$            & $21.56 \pm 0.05$ mas &                    \\
$\phi_2$            & 0 (fixed)            &                    \\
$\chi^2_\mathrm{r}$ & $2.6$                &                    \\
\hline
\end{tabular}
\label{tab:binary}
\end{table}

\begin{figure}
  \resizebox{\hsize}{!}{\includegraphics{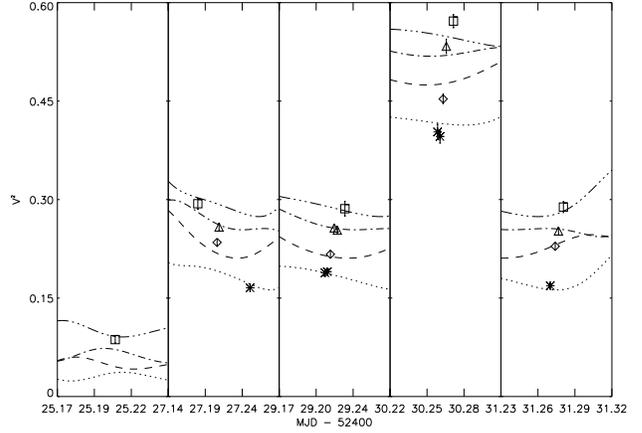}}
  \caption{
  Squared visibility as a function of time, and the best fitting
  binary model.
  }
  \label{fig:v2_mjd}
\end{figure}

\begin{figure}
  \resizebox{\hsize}{!}{\includegraphics{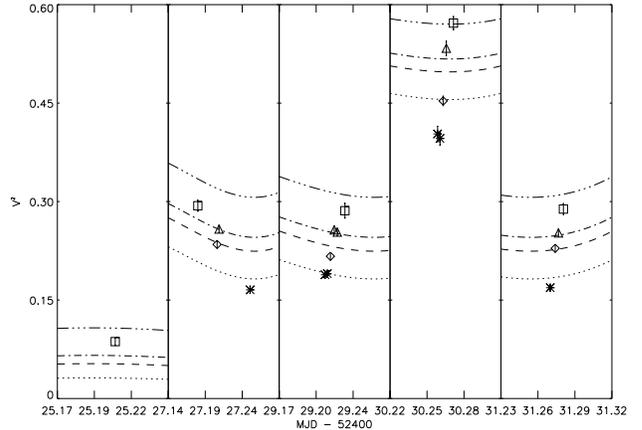}}
  \caption{
  Squared visibility as a function of time, and the best fitting
  single-star SPH model.
  }
  \label{fig:ld_mjd}
\end{figure}

\begin{figure}
  \resizebox{\hsize}{!}{\includegraphics{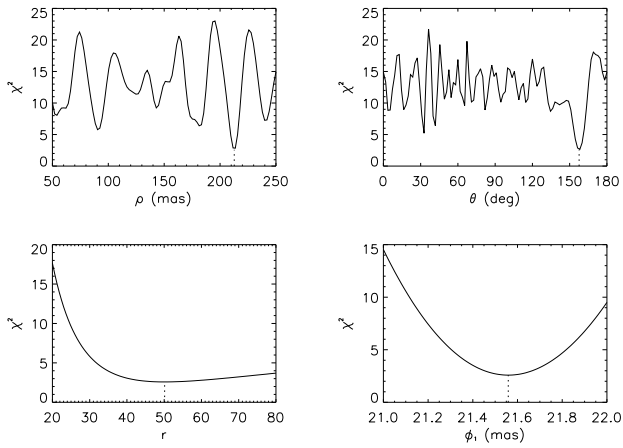}}
  \caption{
  Marginal $\chi^2$ for separation $\rho$, position angle $\theta$,
  contrast ratio $r$ and primary angular diameter $\phi_1$. Global
  minima are marked by a dotted line.
  }
  \label{fig:chi2_marginal}
\end{figure}

\begin{figure}
  \resizebox{\hsize}{!}{\includegraphics{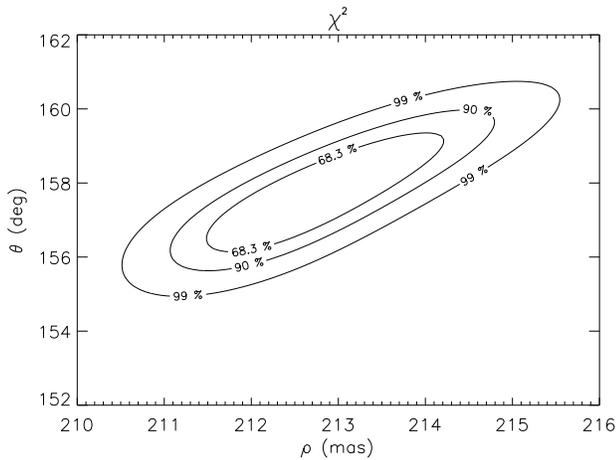}}
  \caption{
  Confidence intervals as a function of separation $\rho$ and position
  angle $\theta$.
  }
  \label{fig:sep_pa}
\end{figure}

\begin{figure}
  \resizebox{\hsize}{!}{\includegraphics{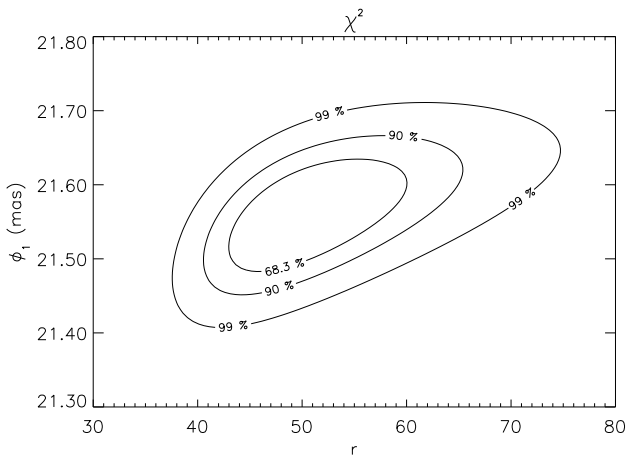}}
  \caption{
  Confidence intervals as a function of contrast ratio $r$ and primary
  angular diameter $\phi_1$.
  }
  \label{fig:r_phi1}
\end{figure}

The best solution (Table~\ref{tab:binary}) has a reduced $\chi^2$ of 2.6,
twice as low as the best single-star model.
For this solution, we have plotted in Fig.~\ref{fig:v2_mjd} the expected
evolution of the squared visibility during our five nights on Arcturus together
with the measured values. For comparison, Fig.~\ref{fig:ld_mjd} presents the
same thing for the single-star SPH model. From the marginal $\chi^2_\mathrm{r}$
curves in Fig.~\ref{fig:chi2_marginal}, we see that there is no ambiguity on the
global minimum with respect to $r$ and $\phi_1$, the situation being more delicate
for $\rho$ and $\theta$ owing to the trigonometric functions in the model.
Figures~\ref{fig:sep_pa} and \ref{fig:r_phi1} show the confidence regions for
the best solution as contours of the $\chi^2$ surface in the two subspaces
$(\rho,\theta)$ and $(r,\phi_1)$. The form of Eq.~\ref{eq:v2_bin} favours the
correlation between the parameters in each of these pairs (98\,\% and 57\,\%
respectively), something that can be seen from the elongated shapes of the
confidence regions as well.
%
%_________________________________________ Fig. V2=f(MJD)
%
%
%_________________________________________ Fig. V2_LD=f(MJD)
%
%
%_________________________________________ Fig. marginal chi2 curves
%
%
%_________________________________________ Fig. chi2=f(sep,pa)
%
%
%_________________________________________ Fig. chi2=f(r,diam1)
%

\subsection{Discussion}

Although Arcturus is generally believed to be a single star, a
literature study reveals that {\sl Hipparcos} also lists Arcturus as a
double star. This result caused some debate at that time
\citep[e.g.][]{Griffin1998, Turner1999} and in response, the {\sl
Hipparcos} team re-analyzed the data, and -- though they arrived at
the same conclusion -- flagged the solution as ``unreliable''
\citep{Soderhjelm}.  Surprisingly, our solution agrees quite well with
the {\sl Hipparcos} solution (Table~\ref{tab:binary}) and we are
therefore convinced that this hypothesis deserves further
investigation. In the following, we will first rule out the
possibility of a background star and a double calibrator, then we
review briefly all relevant published works on Arcturus and finally
confront these with our new results.

\subsubsection{Background star}

As a first step, we can rule out the possibility of a visual binary,
in the sense of a purely optical effect, since the high proper motion
of Arcturus \citep[$-1093.43$ and $-1999.43$\,mas/yr in $\alpha$
resp. $\delta$,][]{Hipparcos} would have increased the separation with
the hypothetical background star by more than $10\arcsec$ since the
time of the {\sl Hipparcos} observations, clearly separating the
fringe packets of both stars. Furthermore, we obtained a new optical
image of Arcturus and its neighborhood using the new CCD camera
MEROPE on the 1.2\,m Belgian Mercator telescope
(http://www.mercator.iac.es/instruments). Images were made in all
filters of the Geneva system \citep[e.g.][]{RufNic} but only in the
(narrow) U band and with an exposure time below 0.01\,s did the
flux not reach the saturation limit of the camera. This image
confirmed that no other star of sufficient magnitude is present at
Arcturus' location in May 2002 (about $5\arcsec$ north-east from its
current location) and at the time of the {\sl Hipparcos} observations.

\subsubsection{Binary calibrator}
\label{sec:bin_cal}

 If the calibrator HR\,5512 were a close binary with a sufficiently
bright companion, using a UD model to compute the interferometric
efficiency might introduce the calibrator binary signature into the
science observations.  However, in Sect.~\ref{sec:obs} we already
argued that HR\,5512 is not a known binary and in
Sect.~\ref{sec:atmos_discussion} we showed how fitting the 2001 data
set (for which other calibrator stars were used) with a LD model yielded
similar residuals, resulting in a $\chi^2_{\mathrm{r}} = 64.5$.  We
are therefore confident that the effects reported here are not due to
binarity of the calibrator.

Fitting a binary model to this 2001 dataset on Arcturus is possible,
and returns parameters roughly consistent with those determined from
the 2002 dataset: a separation of $188.5 \pm 5.3$\,mas and a magnitude
difference of 3.8\,mag.  The resulting $\chi^2_{\mathrm{r}}$ of 1.65
is a clear improvement over the LD disk model. We must remark that the
2001 position angle (PA $=78\degr$) differs significantly from that of
2002, but with such a limited dataset, this result is not very
meaningful.

\subsubsection{The literature}

\paragraph{Radial velocities:}

The radial velocity of Arcturus has been studied for over a century
now \citep[e.g.][]{Lord1904} and, because of the stability of its
velocity, has more recently become an IAU radial velocity standard
\citep{Pearce1955}.

Nevertheless, velocity variations on several time-scales have been
reported in the last decades. \cite{Irwin89} obtained precise
measurements of the radial velocity on 43 occasions between 1981 and
1985. These show a range of 500\,m/s, with both short-period and
long-period variability. For the long-period variability, they find an
amplitude of 120 to 190\,m/s and a period of 640--690 days. Since this
period is longer than the fundamental radial mode of oscillation, they
discuss other possibilities: convection cells and dark spot are not
really compatible with the absence of line-width changes. A companion
with an $M \sin i$ of 1.5 to 7.0 Jupiter masses could explain
their data.  Explaining the long-period variability with beating of
higher frequencies did not work well. In the last decade, a mode of
pulsation with an even longer period has been discovered: the
gravity-mode. However, the periods associated with the g-mode are
still well below 2 years, and it is not certain that these pulsations
are observable through an extended giant atmosphere (Mazumdar, private
communication).

Although he does not present a detailed analysis of his data in this context,
\cite{Cochran88} also notes a long period variability in his
observations.  Note that solar-like oscillations have recently been observed in
Arcturus \citep{Merline96,Retter03}.

\paragraph{Astrometry:}

No astrometric evidence for binarity was found by {\sl Hipparcos} during its
4 year lifetime. However, \cite{Gontcharov2001} combined astrometric
ground-based catalogues containing epochs later than 1939 and the {\sl Hipparcos}
catalogue, to obtain new proper motions and to detect non-linear
astrometric behaviour. In their catalogue (``Proper Motions of
Fundamental Stars''), Arcturus is listed as being an astrometric
binary. Gontcharov (private communication) confirmed that the
astrometric offsets (the residuals after subtraction of all linear
motions) are significantly higher than what can be expected for a
single star system. Periods of 5 and 20 years appear to be present in
these data.

\paragraph{Direct evidence:}

 The only other direct evidence is the visual detection of a companion
by {\sl Hipparcos}. The published results are listed in
Table\,\ref{tab:binary}. Originally, no relative motion of the system
was detected during the {\sl Hipparcos} lifetime. However, after
re-analysis of the data, a small relative motion of 4--8\arcsec/yr was
detected, though at a PA which is ``almost at right angles'' with the
published value \citep{Soderhjelm}.

All other attempts at a direct detection of a companion around
Arcturus have returned no positive results. These include the
non-detection in the H$\alpha$ filter with the AO system on the Mount
Wilson 100 Inch Telescope by \cite{Turner1999}. However, looking at
Fig.~1 of this paper, a companion at a 255~mas separation would be
located on the $2.0\times10^3$ contour of the primary's Point Spread
Function (PSF).  Assuming that the PSF peaks at $3.0\times10^4$ at
least, a contrast ratio of 20, i.e. $1.5\times10^3$ at most, would be
below the level of the PSF, making the detection difficult. Moreover,
the reconstructed image is not diffraction-limited (as 65~mas would be
the size of the first Airy ring), and not circularly-symmetric,
showing that residual aberrations are present.  Now speckle noise is
certainly a concern for faint companion detection, so we believe these
observations cannot definitely rule out the companion as found by {\sl
Hipparcos}.

Nevertheless, another non-detection with Keck aperture masking, and
the (apparent) absence of another fringe packet in IOTA/IONIC
(broadband) H-band interferometric observations (Monnier, private
communication) suggest Arcturus to be a single star
indeed. Unfortunately, these data/results were not conclusive
(Monnier, private communication).  \cite{Quirrenbach96} found a
good agreement between their optical interferometric data on Arcturus
and theoretical limb-darkening profiles. Especially the very low
visibility measured around the first null (their Fig.~3c) seems
incompatible with an unresolved companion only 25 times fainter than
Arcturus.

\paragraph{Spectrophotometric observations}
\label{sec:uv}
It is noteworthy here that recent NLTE atmospheric modelling and
spectrum synthesis calculations have been unable to match the observed
spectrophotometric data in the violet and near UV bands: the flux
observed between 300 and 400~nm amounts to only half of the predicted
flux \citep{Short2003}. While this unresolved discrepancy might have
nothing to do with binarity, it is a further clue that there may be
more to Arcturus than is known.

\subsubsection{Nature of the companion}
\label{sec:nature_companion}
\begin{figure}
\resizebox{\hsize}{!}{\includegraphics{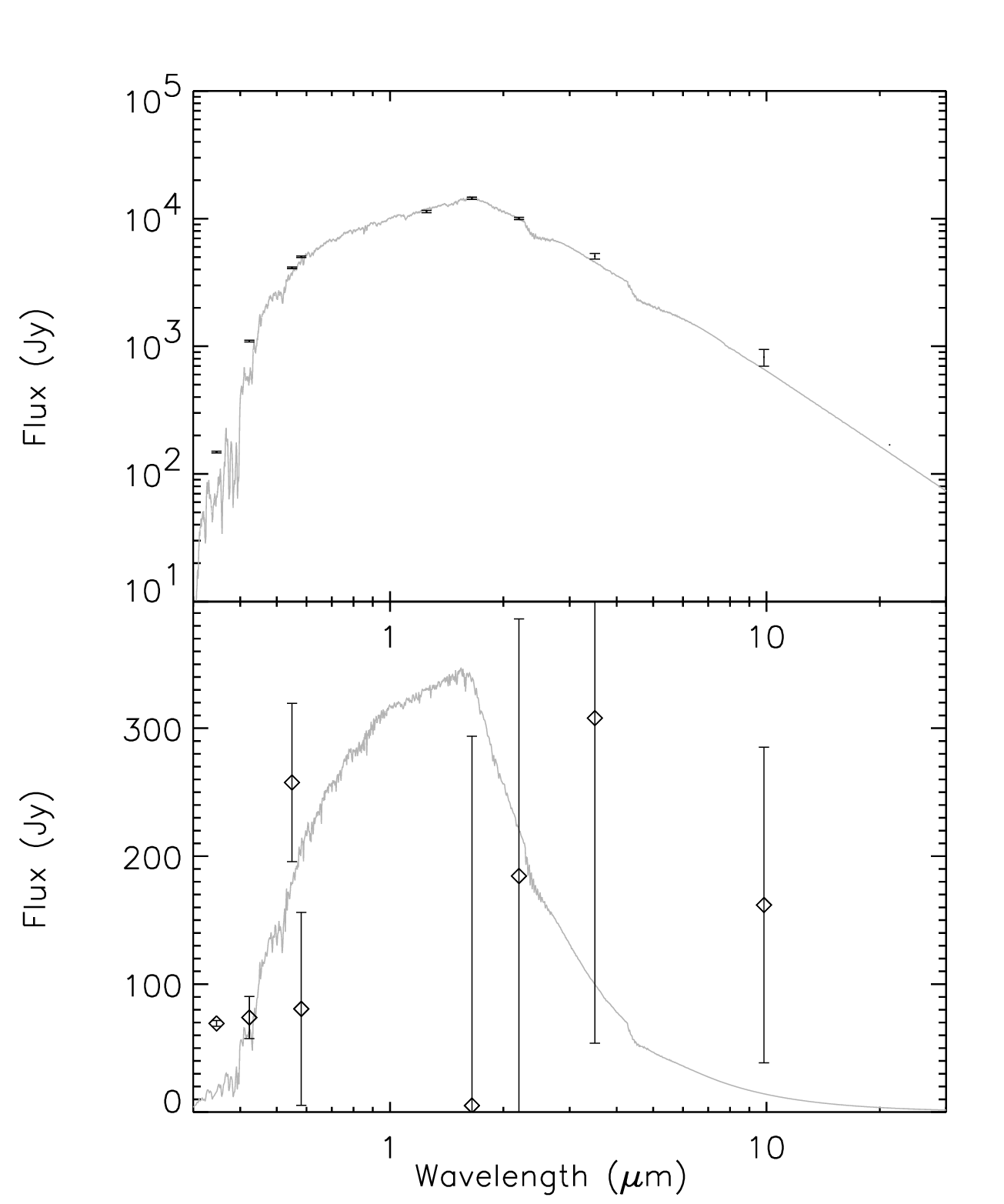}}
  \caption{
  Upper panel: Comparison of new UBVG (Geneva), JHKL (SAAO) and IRAS
  fluxes with a synthetic SED computed from our {\sc marcs} model
  atmosphere for Arcturus (A). There is no clear indication of a
  composite SED. Lower panel: the companion we suggest here
  (Kurucz model shown), would not be detectable in the residuals after
  subtraction of the primary.
}
  \label{SED}
\end{figure}

According to our binary model (Table~\ref{tab:binary}), the contrast
ratio primary/companion in the K band amounts to about 50.
 The UV problem mentioned in Sect.~\ref{sec:uv} not taken into account,
there is no clear photometric evidence of a composite SED/spectrum
(Fig.~\ref{SED}). Therefore, the companion should emit as an object of
fairly similar spectral type: if it were significantly warmer, the
contrast ratio in the optical would be much lower, resulting in a
clearly composite SED. If the companion were cooler (but coeval), it
would have to be a giant further in its evolution and thus more
massive and brighter than the primary\footnote{
Note that a Brown Dwarf is not a possibility for it would not
reach the required brightness (4~mag) in the K band as derived from the
contrast ratio $r$.
}
. This is clearly not a
possibility.  We can therefore conclude that this contrast ratio is a
good upper limit on the luminosity ratio of the two stars. In case of
a slightly warmer companion, the luminosity ratio could possibly be a
factor of two lower, i.e. 25. Adopting a total luminosity of
$196(\pm21)\,\rm{L}_{\odot}$ as determined by \cite{Decin2000}, and a
luminosity ratio range of 25--50, the primary would contribute
188--192\,L$_{\odot}$ and the companion 4--8\,L$_{\odot}$.

A companion of similar spectral type and 4--8\,L$_{\odot}$ can only be
a red (sub-)\-giant of slightly lower initial mass than Arcturus. Note that a
higher mass is not possible because of evolutionary reasons: in such
case, the companion would have been much brighter than the primary,
under the assumption that they are coeval.

\begin{figure}
\resizebox{\hsize}{!}{\includegraphics{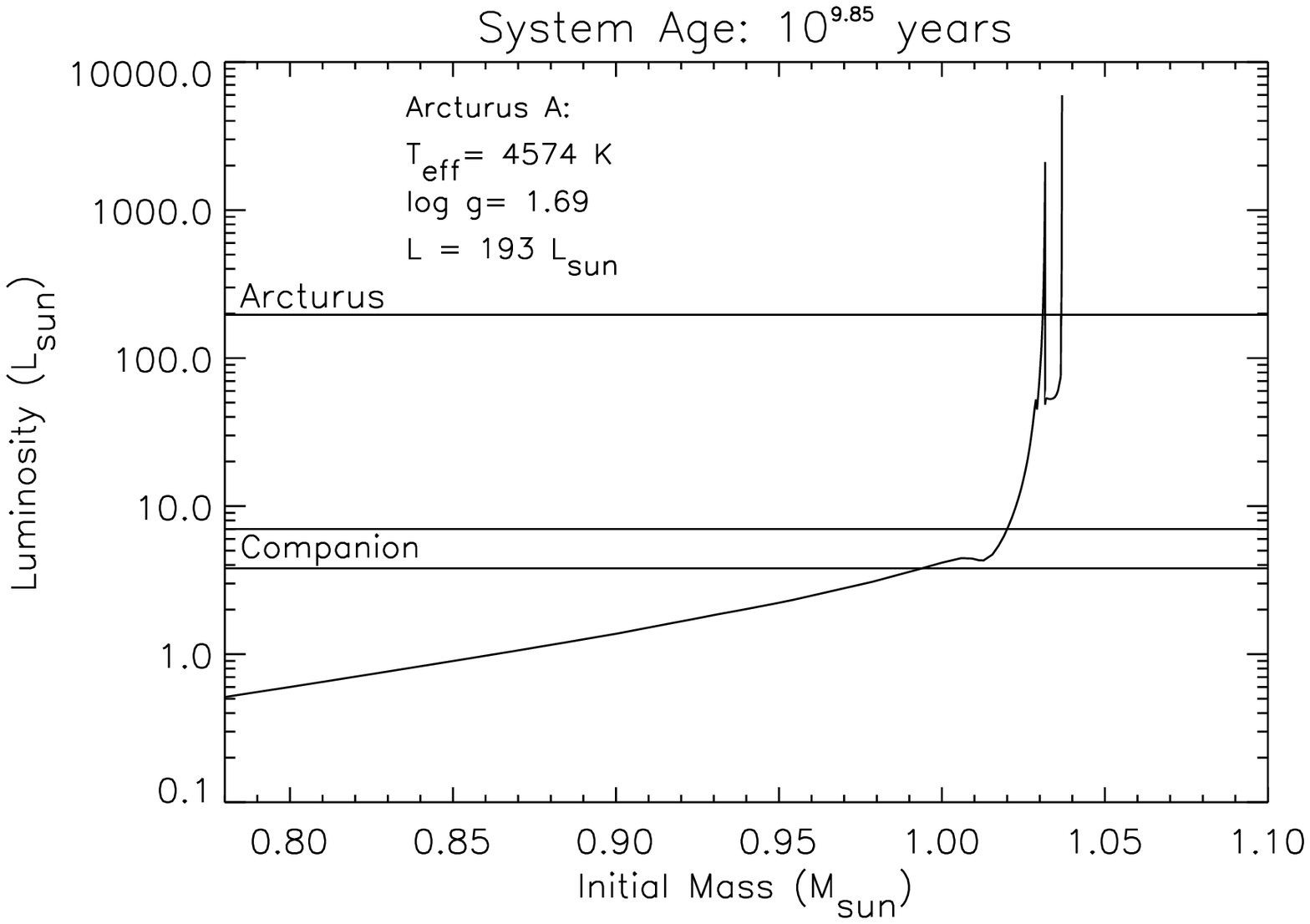}}
\resizebox{\hsize}{!}{\includegraphics{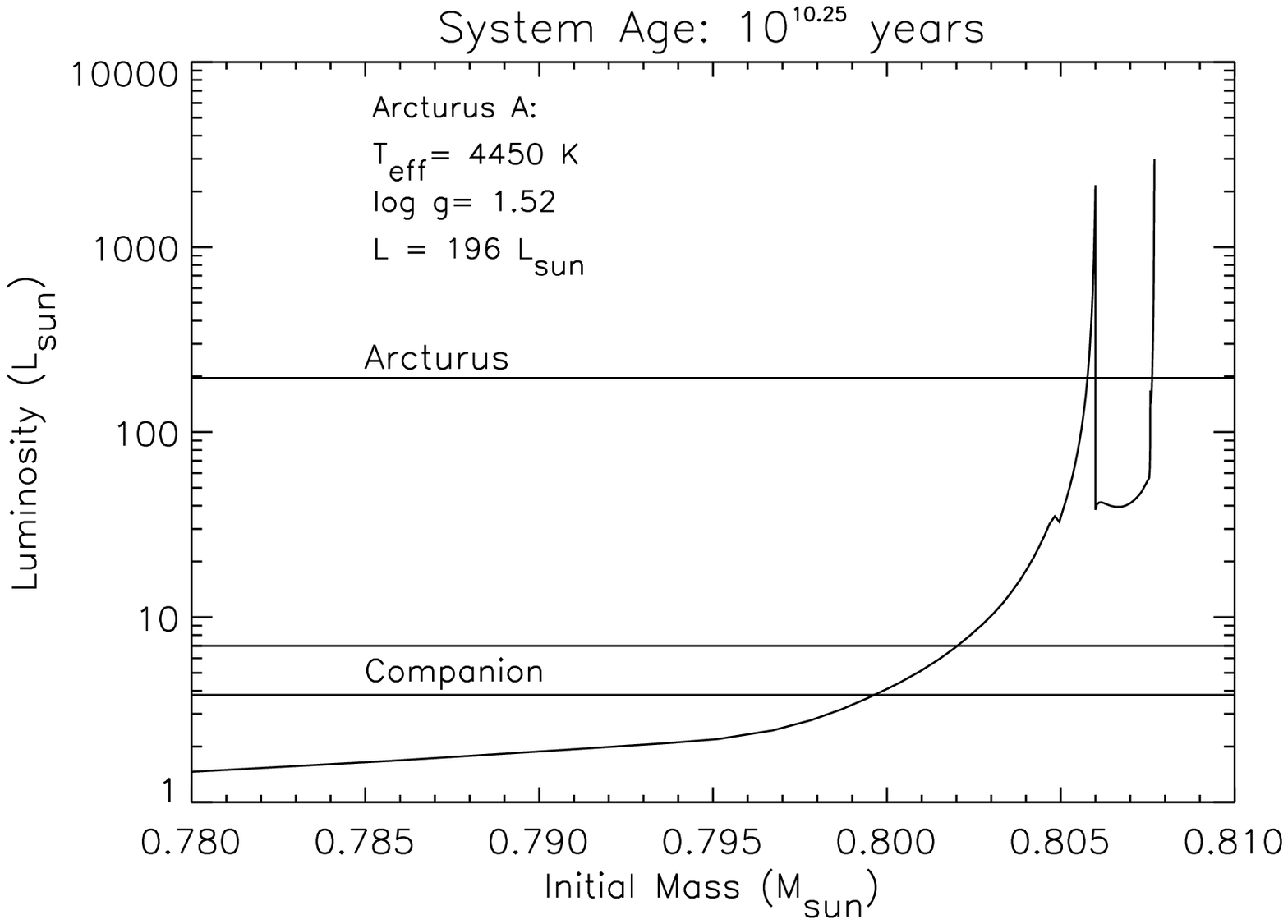}}
  \caption{
  Luminosity as a function of initial mass for a system of $10^{9.85}$
  and $10^{10.25}$ years old. For the companion, we indicate the
  possible range in luminosity The two peaks in luminosity on the
  right hand side correspond to the RGB and AGB phase respectively.
  }
  \label{SA}
\end{figure}

A quantitative estimate of the allowed mass ratio and spectral types
can be made using synthetical isochrones/evolutionary tracks. Using
the Padova database of stellar evolutionary tracks \citep{Girardi} for
$z=0.004$, and the luminosity, $T_\mathrm{eff}$ and $\log{g}$ of
\cite{Decin2000}, we find for the primary (and the possible system) an
age range from about $10^{9.85}$ to $10^{10.25}$\,years (7 to 17 billion
years, actually limited by the age of the universe). In Fig.~\ref{SA},
we present luminosity as a function of initial mass for the lower and
upper limit on the age of the system. 

\begin{table}[ht]
\begin{center}
\caption{
 Primary and companion stellar parameters for age-limits$^2$.\label{age_table}
 }
\vspace{2ex}
\begin{tabular}{l|r|r|r|r}
\hline
\hline
Log(Age)        & \multicolumn{2}{c|}{$9.85$} &
                     \multicolumn{2}{c}{$10.25$}            \\
Component          & A     & B           & A       & B           \\
\hline
L (L$_{\odot}$)    & 195   & 3.8--8      & 195     & 3.8--8      \\
T$_{\rm{eff}}$ (K) & 4574  & 5050--6300  & 4450  & 4950--5050  \\
$\log g$              & 1.69  & 3.5--4      & 1.52    & 3.5         \\
M$_{\rm{initial}}$ & 1.03  & 0.99--1.02  & 0.805   & 0.799--0.802 \\
M$_{\rm{actual}}$  & 1.02  & 0.99--1.01  & 0.796   & 0.796        \\
Spect. Type        & K0III & G5IV-F7IV   & K1.5III & G5IV         \\
\hline
\end{tabular}
\end{center}
\end{table}

In Table~\ref{age_table}, we present the stellar parameters for
primary and companion at the two edges of the age
interval\footnote{The actual T$_{\rm{eff}}$ determined by
\cite{Decin2000} is beyond the grid-limit, but by extrapolation one
can see that this probably does not influence our results.}. From
this table, it is clear that the stellar parameters determined by
\cite{Decin2000} correspond to a fairly old system, but the values
found for the 7-billion-year system are still compatible with the
literature on $\alpha$\,Bootis. Note that the lower mass in the
``old'' solution corresponds to the mass estimate implied by the
measured radius and the derived gravity, which is close to
0.8\,M$_{\odot}$.  Furthermore, the V$-$K colour (1.98) one obtains for
the companion when combining our K-band solution with the {\sl
Hipparcos} V-band solution corresponds to a spectral type of G4IV
\citep[e.g.][]{BCP}, exactly what can be expected from
Table~\ref{age_table} and our preference for a high-age solution! To
obtain a K-band contrast of 50, the companion would need to be about
2.7\,mas large. This value is entirely compatible with the
evolutionary model of the companion put at a distance of $11.26 \pm
0.09$\,pc (derived from the parallax) and with the diameter computed
with the empirical surface brightness relation from
\cite{Kervella2004}, which yields $2.72\pm0.04$. It is compatible as well with
the binary parameters reported in Table~\ref{tab:binary}, as
recomputing the binary model with $\phi_2 = 2.7$~mas does not lead to
a noticeable change in the solution.

In the lower panel of Fig.~\ref{SED}, we present the absolute flux of
this hypothetical companion together with the residuals we obtain by
subtracting a model for the primary from the observed
photometry\footnote{The diameter of the primary was chosen in such a
way as to obtain positive residuals (except for the J-band which is
notoriously sensitive to the water column density in the earth's
atmosphere). This required a slightly smaller diameter than derived
in Sect.~\ref{sec:diameters}, i.e. 20.0\,mas.}. Clearly, the companion
would not be detectable in these residuals, except maybe for the B and
V bands (U should be treated with much care because it is highly
surface-gravity sensitive), which show significant excess, compatible
with the suggested 2.7-mas G4IV companion. 

Note that we found the current secondary mass to be almost equal to
the primary mass. A mass ratio so close to unity, unlikely as it
seems, is compatible with statistical binary mass ratio studies
showing a bimodal distribution with one peak toward a ratio of 1
\citep[e.g.][]{trimble}. The difference in luminosity is the result of
the companion being slightly behind in evolution.

\subsubsection{The orbit}

Combining all of these results, two plausible types of orbit remain: (1)
a very narrow system (separation of just a few AU) seen face-on or (2)
a very wide system (period of the order of centuries, or larger) seen
nearly edge-on.

The first solution does require us to discard the
PA/relative motion detected in the {\sl Hipparcos} data, which is
acceptable given the doubts expressed by \cite{Soderhjelm} in the
re-analysis of the data. It would however be compatible with the
periods found in radial velocities and astrometry of the order of a
few years. To explain the low observed radial
velocity variations, the inclination must be very close to face-on, or
the companion must have a mass of just a few Jupiter masses, which is
not compatible with the observed luminosity.

The second solution would require a nearly edge-on disk to produce
the small projected separation, and the long orbital period would
explain the non-detection of high-amplitude radial velocity and
astrometric variability.  Clearly, it is not possible to combine all
results into one consistent explanation.

%%%%%%%%%%%%%%%%%%%%%%%%%%%%%%%%%%%%%%%%%%%%%%%%%%%%%%%%%%%%%%%%%%%%%

\section{Conclusions and Outlook}
\label{sec:conclusions}

We have presented a new set of narrow-band near-IR interferometric
observations of the K2 giant Arcturus, obtained to test the
applicability of K giants and K giant models to the calibration of
high-accuracy infrared interferometric observations. A comparison with
state-of-the-art stellar atmospheres failed to explain the data: the
residuals show clear systematics. These are independent of calibration
and none of our hypotheses about the stellar atmosphere (extension, spots) can
explain the data. A binary model, with a sub-giant as companion, does
provide a good fit to the data. A thorough
literature study reveals that there is ample, but inconclusive and
inconsistent evidence for a companion. 

\subsection{Implications on calibration}
If the calibrator one uses is an unknown binary, or
the effects reported in this paper are due to some other unknown
characteristics of K giant stars, then the error on the visibility
could be increased by a factor as large as $\sqrt{\chi^2_\mathrm{r}} = \sqrt{5.6}
\approx 2.5$, if a standard model atmosphere is used to represent the
calibrator. For single-mode interferometers such as FLUOR or VINCI,
it means that the standard error on the visbility would jump from 1--2\,\%
to 2.5--5\,\%. Further investigation is urgently required to settle
this major issue. This should
include both new observations of other ``normal'' K giants to check
the uniqueness of the problems reported here, and the re-observation
of Arcturus to pin-point with more certitude the nature of this star.

\subsection{Prospects for Arcturus}
Unfortunately, our dataset which was aimed at determining the
wavelength dependence of the diameter of Arcturus (A), does not allow
a full characterization of the (still hypothetical) system. New
attempts at imaging with the most recent AO systems might resolve the
system, though the brightness of the primary will seriously complicate
the observations. Another possibility would be a new interferometric
dataset with two telescopes on similar or shorter baselines, but
obtained over a wide range of azimuth angles, such that the companion
would pass through several of the fringes, providing a much clearer
signal than was possible with the present essentially snapshot data
set.  Yet another possibility would be an interferometric data set
with three telescopes (e.g. the present IOTA), using two short
baselines to lock on the fringes, and one long baseline on which the
large-diameter primary would have weak fringes but the small-diameter
secondary would have relatively strong fringes, again observing
over a large range of position angles so that the secondary passes
through many fringes, for a clear signature.

%%%%%%%%%%%%%%%%%%%%%%%%%%%%%%%%%%%%%%%%%%%%%%%%%%%%%%%%%%%%%%%%%%%%%%

\begin{acknowledgements}

The authors would like to thank John Monnier for his search of a faint
companion in his Keck aperture masking and IOTA/IONIC observations of
Arcturus, as well as Guillermo Torres for his valuable comments on a
draft of this paper and the referee, dr. Ian Short, for his critical
reading and suggestions.  T.V. appreciated support by the European
Community through a Marie Curie Training Fellowship for an extended
stay at Paris-Meudon Observatory. (The European Community is not
responsible for the information communicated.)  This work was also
performed in part under contract with the Jet Propulsion Laboratory
(JPL) funded by NASA through the Michelson Fellowship Program. JPL is
managed for NASA by the California Institute of Technology.

\end{acknowledgements}

\bibliographystyle{aa}
\bibliography{references}

\end{document}